\documentclass[twocolumn,prl,aps,superscriptaddress,showpacs]{revtex4-1}

\usepackage{lineno}
  \usepackage{mathptmx}
\usepackage{subfigure}
\usepackage{dcolumn}
\usepackage{amsmath,amssymb}
\usepackage{bm}
\usepackage{color}
\usepackage{overpic}
\usepackage{latexsym}
\usepackage{epstopdf}
\usepackage{color}
\usepackage[english]{babel}
\usepackage{latexsym}
\usepackage{stmaryrd}

\usepackage{psfrag,graphicx} 
\usepackage{epsf} 
\usepackage{subfigure} 
\usepackage{amsmath} 
\usepackage{amssymb} 
\usepackage{amsfonts}
\usepackage{bm}
\usepackage{natbib}
\usepackage{epstopdf}
\usepackage{appendix}

\definecolor{mygrey}{gray}{0.35}
\definecolor{myblue}{rgb}{0.2,0.2,0.8}
\definecolor{myzard}{cmyk}{0,0,0.05,0}
\definecolor{mywhite}{rgb}{1,1,1}
\definecolor{myred}{rgb}{1,0.,0.3}

\usepackage[colorlinks=true,citecolor=myblue,linkcolor=myred]{hyperref}
\def\be{\begin{equation}}
\def\ee{\end{equation}}
\def\ba{\begin{align}}
\def\enda{\end{align}}
\def\bi{\begin{itemize}}
\def\ei{\end{itemize}}

 \def\ee{\mathord{\rm e}}
 
 \def\ii{\mathord{\rm i}}

 \def\ee{\mathord{\rm e}}
 
 \def\ii{\mathord{\rm i}}

\renewcommand{\ii}{{\rm i}}
\renewcommand{\ee}{{\rm e}}

\def\beq{\begin{equation}}
\def\beq{\begin{equation}}
\def\eeq{\end{equation}}

 \newcommand{\ket}[1]{|#1\rangle}

\begin{document}


\title[Short Title]{Frustrated Quantum Spin Models with Cold Coulomb Crystals}
\author{A. Bermudez}
\affiliation{Institut f\"{u}r Theoretische Physik, Albert-Einstein Allee 11, Universit\"{a}t Ulm, 89069 Ulm, Germany}
\author{J. Almeida}
\affiliation{Institut f\"{u}r Theoretische Physik, Albert-Einstein Allee 11, Universit\"{a}t Ulm, 89069 Ulm, Germany}
\author{F. Schmidt-Kaler}
\affiliation{Institut f\"{u}r Physik, Staudingerweg 7, Johannes Gutenberg-Universit\"{a}t Mainz, 55099 Mainz, Germany}
\author{A. Retzker}
\affiliation{Institut f\"{u}r Theoretische Physik, Albert-Einstein Allee 11, Universit\"{a}t Ulm, 89069 Ulm, Germany}
\author{M. B. Plenio}
\affiliation{Institut f\"{u}r Theoretische Physik, Albert-Einstein Allee 11, Universit\"{a}t Ulm, 89069 Ulm, Germany}

\begin{abstract}
We exploit the geometry of a zig-zag cold-ion crystal in a linear trap to propose the quantum simulation of a paradigmatic model of long-ranged magnetic  frustration. Such a quantum simulation would  clarify the complex features of a rich phase diagram that  
presents  ferromagnetic, dimerized antiferromagnetic, paramagnetic, and floating phases, together with previously unnoticed  features that are
hard to assess by numerics. We analyze in detail its experimental feasibility, and provide supporting numerical evidence 
 on the basis of realistic parameters in current ion-trap technology.
\end{abstract}

\maketitle

Our understanding  of interacting quantum many-body  systems is  usually hindered by their inherent complexity. In order  to unveil their properties, an original approach  are the so-called {\it quantum simulations} (QSs), which exploit a well-controlled quantum system to study a complicated interacting model~\cite{qs}. In particular, the QS of {\it magnetic ordering} is a promising direction of research. Although this effect was  already identified by the ancient greeks, it still presents some puzzles that  would benefit from a quantum simulator since they cannot be efficiently addressed numerically~\cite{spin_liquids}. Among the most prominent  platforms for the QS of magnetism, such as neutral atoms or photons~\cite{qs_magnetism},  trapped ions~\cite{ising_ions} have the advantage of near unity fidelity in state preparation, individual readout, and a wide tunability of interactions.  Here, we investigate the QS of {\it frustrated  magnetism} with trapped ions, a phenomenon that yields a fundamental challenge  with important connections to high-temperature superconductivity and other exotic states of matter, such as spin liquids/glasses~\cite{spin_liquids}.

Even if the typical distances between trapped ions  make a direct dipole-dipole coupling negligible; the  long-range Coulomb interaction yields  an indirect mechanism whereby the vibrational phonons, coupled to the spins by lasers, mediate a spin coupling~\cite{ising_porras}. In the seminal work~\cite{frustration_monroe}, frustration in a 3-ion chain has been achieved by fine tuning the laser frequencies within the collective vibrational band. However, scaling  this scheme to larger ion chains may find a fundamental limit:  the vibrational band becomes increasingly  dense, and the  lasers approach a certain resonance. An  alternative  scheme would use micro-fabricated  surface traps~\cite{surface_traps}, but the current  ion-heating problems must  be overcome  to obtain strong-enough  interactions. One may instead rely  on  large trapping non-linearities~\cite{frustrated_hard_core_bosons_schmied}, yet the requirements remain to be met in experiments. Here, we present a different method  amenable of being scaled to  larger  ion numbers that relies on state-of-the-art ion traps and spin-dependent  light forces~\cite{sd_force}. 

 \begin{figure}

\centering
\includegraphics[width=1\columnwidth]{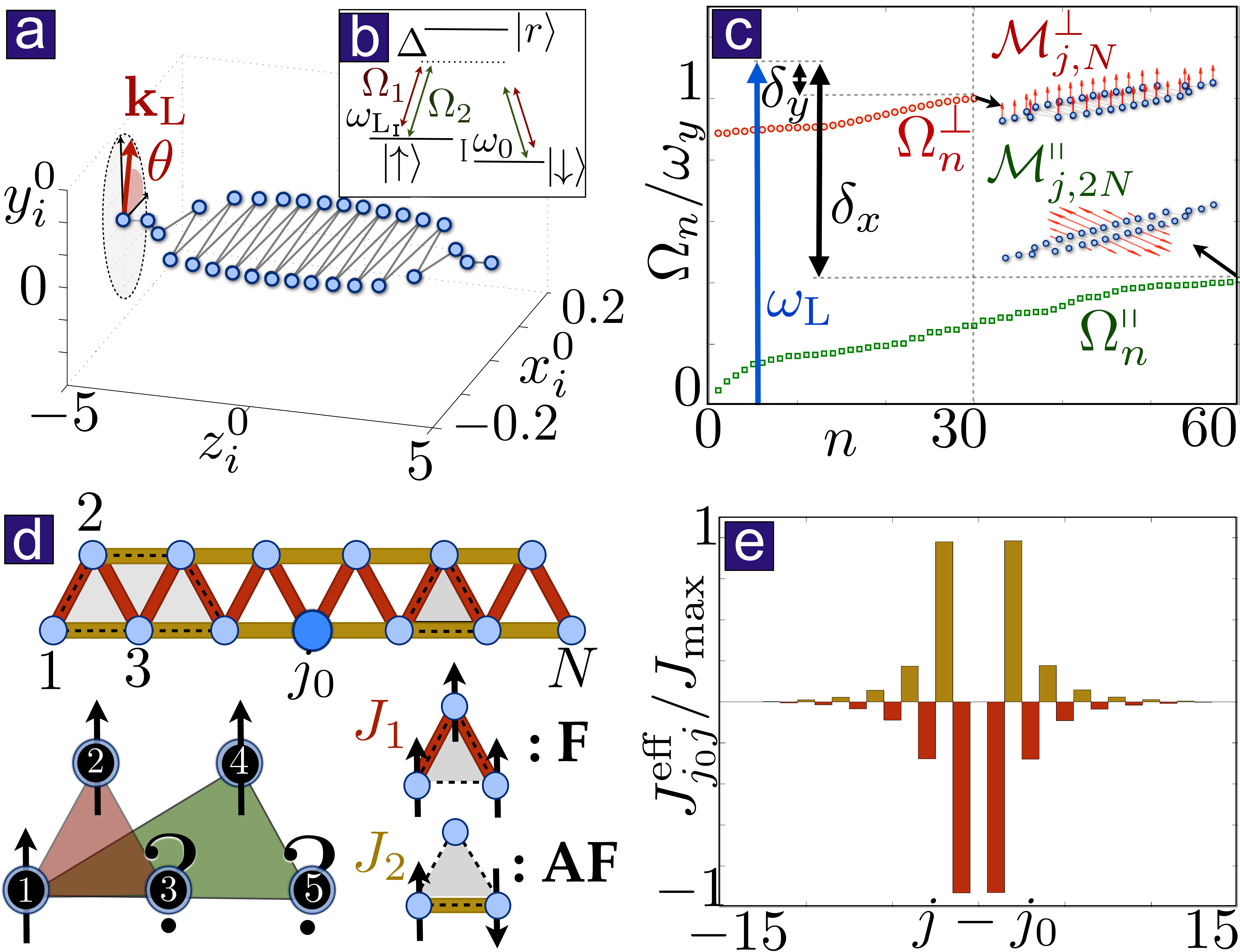}
\caption{{\bf Zig-zag ladder:} {\bf(a)} Equilibrium positions in units of $l_z$  for  $N=30$ ions, $\omega_y\gg \omega_x=6.1\omega_z$, as given by $\partial (V_{\rm trap}+V_{\rm Coul})/\partial r_{j\alpha}=0$. {\bf(b)} Scheme of the stimulated Raman transition. A pair of lasers with  frequencies $\omega_1,\omega_2$, and wavevectors ${\bf k}_1,{\bf k}_2$, are far-detuned from  the transitions and  lead to a two-photon transition  $\omega_{\rm L}=\omega_1-\omega_2$. {\bf(c)} Transverse, $\Omega_n^{\bot}$, and planar, $\Omega_n^{\shortparallel}$, normal-mode frequencies. In the inset, the amplitudes of the highest-frequency modes. We also show the frequency $\omega_{\rm L}$ of a  laser beam blue-detuned from the vibrational band leading to  the transverse  spin-dependent force. {\bf (d)} Scheme of the spin  interactions, where $J_1<0$ ($J_2>0$) is  a ferromagnetic (antiferromagnetic) coupling.  Incompatible frustration of two  plaquettes due to the dipolar range. {\bf (e)} Coupling  between  the central ion $j_0=N/2$ and its neighbors, as obtained for $k_{{\rm L},x} l_z=\pi/(x_{j_0}^0-x_{j_0+1}^0)$.}
\label{ladder_scheme}
\end{figure}

 Trapped  ions crystallize  in  self-organized Coulomb crystals at low temperatures.  By tuning the trapping frequencies~\cite{structural_transition},  a linear string of ions undergoes a  structural transition   to a {\it two-leg  zig-zag ladder} [Fig.~\ref{ladder_scheme}{\bf (a)}]. Whereas the linear configuration has been the usual scenario for the QS of magnetism~\cite{ising_ions}, the capabilities of the zig-zag ladder have remained largely unexplored. In this Letter, we fill in such gap. We show that
by exploiting  the commensurability of the inter-leg distance of the ladder with the  wavelength of the spin-dependent light force, it is possible to control both the relative strength and  the sign of the spin interactions in a way that is almost independent of the crystal size, and can be thus scaled to larger ion crystals.  We describe how to reach regimes of competing  ferromagnetic ({\bf F}) and antiferromagnetic ({\bf AF}) couplings, which are incompatible  with the  crystal geometry so that the groundstate cannot minimize all interactions simultaneously. Therefore, trapped ions can be used as quantum simulators of the two main sources of frustration, namely, {\it geometric frustration} and {\it competing interactions}.  

Let us summarize our main results: {\it (i)} The sign of the spin-spin interactions between distant ions depends on wether they lie along the same or different rungs of the ladder.  {\it (ii)} The relative strength of these two  interactions can be controlled by tuning the ratio of the inter-leg distance to the wavelength of the light forces. {\it (iii)} These ingredients yield a fully-tunable and experimentally-feasible quantum simulator  of the   {\it  $J_1$-$J_2$ quantum Ising model} ($J_1$-$J_2$-$QIM$),  a cornerstone of magnetic frustration~\cite{classical_anni_review}. {\it (iv)} The long-range character of the spin couplings introduce incompatible sources of frustration leading to a richer phase diagram with a novel conjectured phase.

{\it Model.--} We consider  $N$ ions of mass $m$ and charge $e$, confined in a linear trap with  frequencies $\{\omega_{\alpha}\}_{\alpha=x,y,z}$~\cite{wineland_review}.  The geometry of  the Coulomb crystal ${\bf r}_j^0$,  determined by the balance of the trapping forces and the Coulomb repulsion, corresponds to a zig-zag ladder in the $x$-$z$ plane for $\omega_z< \omega_x\ll \omega_y$~\cite{structural_transition}. The  vibrations around the equilibrium ${\bf r}_j={\bf r}_j^{0}+ \Delta{\bf r}_j$ are coupled  through the Coulomb interaction, which yields model of coupled oscillators  in the harmonic approximation
\begin{equation}
\label{vib}
H=\sum_{j\alpha}\left(\frac{1}{2m}{ p}_{j\alpha}^2+\frac{1}{2}m{\omega}_{\alpha}^2\Delta r_{j\alpha}^2\hspace{-0.5ex}\right)+\frac{1}{2}\hspace{-0.5ex}\sum_{j k,\alpha\beta}\hspace{-1ex}\mathcal{V}_{jk}^{\alpha\beta}\Delta r_{j\alpha}\Delta r_{k\beta},
\end{equation}
where $\mathcal{V}_{jk}^{\alpha\beta}$ are the vibrational couplings~\cite{comment}. We identify two types of collective excitations, namely  {\it transverse}  and {\it planar} modes that describe the motion perpendicular and parallel to the crystal in terms of the  phonon  operators  $a_n^{\phantom \dagger},b_n^{\phantom \dagger}$. The  vibration amplitudes $\mathcal{M}_{j,n}^{\bot},\mathcal{M}_{j,n}^{\shortparallel},$ and frequencies $\Omega_{n}^{\bot},\Omega_{n}^{\shortparallel},$ are obtained by solving the quadratic problem~\eqref{vib}, leading to
$H_{\rm ph}=\sum_{n=1}^{N}\Omega_n^{\bot}a_n^{\dagger}a_n^{\phantom \dagger}+\sum_{n=1}^{2N}\Omega_{n}^{\shortparallel}b_{n}^{\dagger}b_n^{\phantom \dagger}$ with  the phonon branches displayed in Fig.~\ref{ladder_scheme}{\bf(c)}. As shown in this work, when the planar bandwidth is small enough, a blue-detuned laser beam will only excite the transverse phonons even if it also has  a component along the crystal [Fig.~\ref{ladder_scheme}{\bf(a)}]. This is  a crucial point of our scheme. The transverse phonons, which are best suited to act as carriers of the magnetic interaction,  will be responsible of mediating the spin-spin coupling. The component of the laser along  the zig-zag plane will not excite the planar phonons, but rather open the possibility of  tailoring the ferromagnetic-antiferromagnetic sign of the spin coupling depending on the ratio of the inter-leg ion distance  with the  wavelength of light.

{\it Effective frustrated spin models.--} We consider a pair of laser beams that induce a stimulated Raman transition between two internal states  $\ket{\!\!\uparrow_j},\ket{\!\!\downarrow_j}$ [Fig.~\ref{ladder_scheme}{\bf(b)}]. When the difference between the laser frequencies matches the transverse vibrational frequency,   we get a Stark shift 
$
H_{\rm d}\!=\!\frac{\Omega_{\rm L}}{2}\!\!\sum_j\!\!\sigma_j^z\ee^{\ii{\bf k}_{\rm L}\cdot {\bf r}^0_j}\ee^{\ii({\bf k}_{\rm L}\cdot \Delta{\bf r}_j-\omega_{\rm L} t)}\!\!+\!\!\text{H.c.},
$
where $\sigma_j^z$ is a Pauli matrix, $\Omega_{\rm L}$ is the two-photon Rabi frequency, and ${\bf k}_{\rm L}$ lies in the $x$-$y$ plane. After introducing the phonons, we   make an expansion for small Lamb-Dicke parameters 
$
\eta_{n\bot}=k_{{\rm L,y}}/\sqrt{2m\Omega_{n}^{\bot}}\ll1
$ (similarly for the planar modes).
 By tuning the laser frequency above the transverse phonon branch, we can neglect all non-resonant terms but  the transverse spin-dependent  dipole force
 \begin{equation}
 \label{transverse_pushing}
 H_{\rm d}^{\bot}=\frac{\Omega_{\rm L}}{2}\sum_{j,n}\ii\ee^{\ii{\bf k}_{\rm L}\cdot {\bf r}_j^0}\eta_{n\bot}\mathcal{M}_{jn}^{\bot}\sigma_j^za_{n}^{\dagger}\ee^{\ii\delta_{n\bot}t}+\text{H.c.},
 \end{equation}
 where $\delta_{n\bot}=\Omega_{n}^{\bot}-\omega_{\rm L}$, and we work in a frame rotating with the phonons. For this approximation to hold, the following conditions must be satisfied: $\Omega_{\rm L}\ll \omega_{\rm L}$, $\eta_{n\shortparallel}\Omega_{\rm L}\ll|\Omega_{n}^{\shortparallel}-\omega_{\rm L}|.$
 The first condition is necessary to suppress the carrier terms, and  the second one to avoid the coupling of the spins to the planar modes, and are both satisfied for the realistic parameters below. Additionally, by aligning the lasers nearly parallel to the $y$-axis, we   favor  the fulfillment of the last condition.

  The conditional  force~\eqref{transverse_pushing}, which couples the  spins to the transverse phonons, induces a  spin interaction due to the  virtual exchange of phonons. The exact expression is  obtained by a Lang-Firsov-type transformation~\cite{spin_displacement_wunderlich},\cite{ising_porras} that decouples spins and phonons $\ee^{-S}(H_{\rm ph}+H^{\bot}_{\rm d})\ee^S\approx H_{\rm eff}+H_{\rm ph}$, where
 $S=\frac{\Omega_{\rm L}}{2}\sum_{jn}\ii\ee^{\ii{\bf k}_{\rm L}\cdot{\bf r}_j^0}\frac{\eta_{n\bot}}{\delta_n^{\bot}}\mathcal{M}_{jn}^{\bot}\sigma_j^za_{n}^{\dagger}-\text{H.c.}
$ generates a spin-dependent displacement.  This leads to an effective Ising  model
\begin{equation}
\label{quantum_ising}
\begin{split}
H_{\rm eff}&=\sum_{j\neq k}J_{jk}^{\rm eff}\sigma_j^z\sigma_k^z-h\sum_j\sigma_j^x,\\
J_{jk}^{\rm eff}&=-\sum_{n}\frac{\Omega_{\rm L}^2k_{L}^2\sin^2\theta}{8m\Omega_n^{\bot}\delta_{n\bot}}\mathcal{M}_{jn}^{\bot}\mathcal{M}_{kn}^{\bot}\cos({\bf k}_{\rm L}\cdot {\bf r}_{jk}^0),
\end{split}
\end{equation} 
where the  transverse field, $h$,  follows from a microwave resonant with the transition, and $\sigma_j^x$ is a Pauli matrix. In this work, we are interested in the  anisotropy of $J_{jk}^{\rm eff}\propto\cos({\bf k}_{\rm L}\cdot {\bf r}_{jk}^0)$, which has been  overlooked so far since it vanishes for  linear ion chains. Remarkably, it can lead to magnetic frustration in planar crystals. In the  limit of tight transverse confinement, 
\begin{equation}
\label{ising_strengths}
J_{jk}^{\rm eff}\hspace{-0.2ex}=\hspace{-0.2ex}\frac{J_{\rm eff}\cos\phi_{jk}}{|{\bf r}_{j}^0-{\bf r}_{k}^0|^3},\hspace{0.75ex} J_{\rm eff}\hspace{-0.2ex}=\hspace{-0.2ex}\frac{\Omega_{\rm L}^2\eta_{y}^2\omega_z^2}{8\delta_{y}^2\omega_y},\hspace{0.ex}\phi_{jk}\hspace{-0.2ex}=\hspace{-0.2ex}k_{\rm L,x}l_z(x_j^0-x_k^0)
\end{equation}
where $\eta_{\alpha}=k_{{\rm L},\alpha}/\sqrt{2m\omega_{\alpha}}$, and  $l_z=(e^2/m\omega_z^2)^{1/3}$ gives a unit for the distances. A crucial observation for the QS of frustration is that the sign of the couplings may be  different for ions that belong to the same rung of the ladder, or to distinct ones. More precisely, if the condition $k_{{\rm L},x} l_z=n\pi/(x_j^0-x_{j+1}^0)$ is fulfilled for  a pair of ions in the bulk of the crystal such that $n$ is an integer,  we obtain a ferromagnetic (antiferromagnetic) coupling between pairs of ions in different (same) rungs. We corroborate this claim by obtaining  the exact couplings from the numerical solution of Eq.~\eqref{quantum_ising}, which yields  the aforementioned sign alternation [Fig.~\ref{ladder_scheme}{\bf (e)}] in agreement with the approximate description~\eqref{ising_strengths}. Let us highlight the importance of this result: the groundstate shall not be able to minimize such spin-spin couplings simultaneously, and will thus yield a frustrated magnet~[Fig.~\ref{ladder_scheme}{\bf(d)}]. According to the Tolouse-Villain criterion~\cite{{villain_frustration}}, our spin  model~\eqref{quantum_ising}  is  frustrated since
there is an {\it odd} number of {\bf AF} couplings per unit cell. 
In the following, we support this analytical derivation with numerical simulations on the basis of the realistic experimental settings and parameters that are described now. 

\begin{figure*}

\centering
\includegraphics[width=1.7\columnwidth]{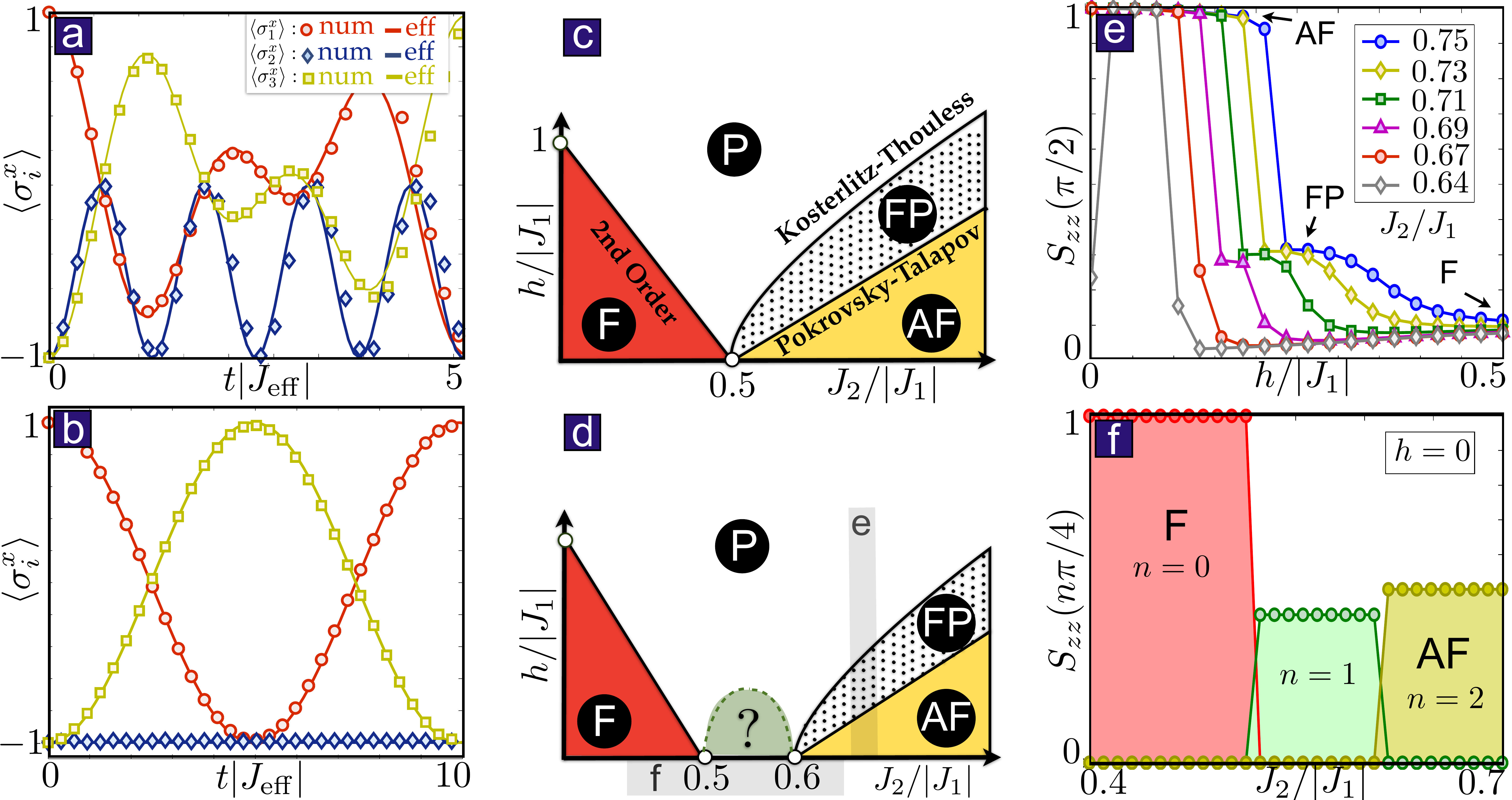}
\caption{ {\bf QS of the frustrated quantum Ising model:}  {\bf (a)} Dynamics of $\langle \sigma_j^x(t)\rangle$ for an initial spin excitation $\ket{+}_1$  in the {\it allowed-hopping} regime. Comparison of the analytical description $H_{\rm eff}$, and  the numerical study of the complete Hamiltonian $H_{\rm ph}+H_{\rm d}$ where the phonon Hilbert space is truncated to one excitation. {\bf (b)} In the {\it inhibited-hopping} regime, there is  a complete agreement between both descriptions, where the spin excitation cannot occupy site $2$. {\bf (c)} Schematic phase diagram of the $J_1$-$J_2$-$QIM$ with  ferromagnetic ({\bf F}), dimerized antiferromagnetic ({\bf AF}), paramagnetic ({\bf P}), and floating ({\bf FP}) phases. {\bf (d)} Conjectured phase diagram with dipolar interactions. Also highlighted in grey the regions studied in {\bf (e)-(f)}. {\bf (e)} $q=\pi/2$ component of the structure factor evidencing the {\bf AF-FP-P} transitions. {\bf (f)} Different magnetic modulations at $h=0$  showing the  splitting of the critical points with a different order in between. (structure-factor components are normalized with respect to the maximum).}
\label{spin_dynamics}
\end{figure*}

 {\it Experimental setup.--} To  realize the  spin model~\eqref{quantum_ising}, we rely on trapped-ion quantum-processing technologies developed to a high degree of perfection~\cite{14_molmer_sorensen}. We emphasize that all procedures and experimental parameters are well in reach with current experimental standards. Cold-ion crystals are kept in linear Paul traps with a combination of static and dynamic electric fields~\cite{wineland_review}. The specific case in Fig.~\ref{ladder_scheme}{\bf(a)} requires a significant anisotropy of the radial confinement $\omega_x,\omega_y$, which can be achieved by deforming the symmetric four-rod trap electrode configuration. Additionally, the anisotropy can be increased and tuned to the desired value by applying a bias voltage to the DC electrodes. In this way, the desired zig-zag crystal is trapped with MHz trap frequencies, and fulfilling the condition $\omega_z < \omega_x \ll \omega_y$. The spins in the generic  level scheme  in Fig.~\ref{ladder_scheme}{\bf (b)} might correspond to the hyperfine states of $^9{\rm Be}^+, ^{43}{\rm Ca}^+$ ions, or to the Zeeman sub-levels of the groundstate of $^{40}{\rm Ca}^+$. These levels are coupled with Raman laser beams that mediate the optical spin-dependent light forces~\cite{sd_force}.

The experimental sequence consists of a number of steps starting with the {\it (i)} initialization of the system. Here, the temperature of all modes is reduced in a Doppler cooling step, followed by a sub-Doppler cooling of the transverse mode. The relatively high frequency $\omega_y/2\pi = 20$MHz ensures good cooling results. The spins are initialized by optical pumping, followed by a $\pi/2$ pulse on a resonant Raman transition (not shown in Fig.~\ref{ladder_scheme}{\bf (b)}) such that a spin superposition is reached. {\it (ii)} The spin-spin interaction is generated by application of the Raman light fields. To suppress spontaneous processes, a large detuning $\Delta >10$GHz is chosen, and light beams with an optical power of about 10mW are focused into a waist of 30 to 60$\mu$m~\cite{schmidt_kahler_cats} suffice to generate $J_{\rm eff} \approx1$kHz. After an interaction time in the millisecond range, {\it (iii)} the simulation result is readout from the fluorescence imaged on the chip of a CCD camera. For this, we illuminate the entire crystal with light resonant to a dipole-allowed transition such that the $\ket{\uparrow}$ state emits fluorescence while $\ket{\downarrow}$ does not. For the  demonstration of frustration, the requirements are much released in comparison to those of  full quantum-state tomography~\cite{haffner_tomography}.

{\it Numerical validation.--} We address the simplest scenario where  frustration occurs: a 3-ion  zig-zag with  trap  frequencies $\omega_y/\omega_z=20,\omega_x/\omega_z=1.43$,  $\omega_z/2\pi\approx1$MHz, leading to 
$
{\bf r}_j^0/l_z\in\{(-0.22,0,-0.92),(0.44,0,0),(-0.22,0,0.92)\}.
$
We test the crucial assumption underlying the derivation of the  frustrated Hamiltonian~\eqref{quantum_ising}, namely the possibility to neglect the pushing force on the planar modes. Even if  the dipole force only acts in the $x$-$y$ plane, the $z$-motion gets coupled through the Coulomb interaction. Hence, we must solve numerically the dynamics of the complete Hamiltonian $H_{\rm ph}+H_{\rm d}$, and compare it to the effective analytical description $H_{\rm eff}$ in~\eqref{quantum_ising}, where we set $h=0$,  $\ket{\psi_0}=\ket{+}_1\otimes\ket{-}_2\otimes\ket{-}_3$, with $\ket{\pm}=(\ket{\uparrow}\pm\ket{\downarrow})/\sqrt{2}$, and consider a ground-state cooled crystal for simplicity. We study two regimes:
{\it (i)}{ Allowed hopping:} We set $\eta_y=0.1=10\eta_x$, $\omega_{\rm L}=1.2\omega_y$, $\Omega_{\rm L}=0.1|\omega_y-\omega_{\rm L}|/\eta_y$, and study the evolution of the whole spin-phonon system for  $k_{\rm L,x}=0$. Since there is no modulation of the coupling sign, the initial spin excitation  located at site 1 hops around the triangular plaquette as a consequence of the Ising coupling. In Fig.~\ref{spin_dynamics}{\bf (a)}, we show the  agreement between the analytical and the numerical descriptions. 
{\it (ii)}{ Inhibited hopping:} We use the same parameters, but set  $k_{\rm L,x}=\pi/2l_z(x^0_2-x^0_1)$. It follows that the Ising couplings between $1$-$2$ and $2$-$3$  vanish $J_{12}^{\rm eff}=J_{23}^{\rm eff}=0$, and thus the excitation  only tunnels along $1\leftrightarrow 3$, as confirmed by both analytical and numerical descriptions in Fig.~\ref{spin_dynamics}{\bf (b)}.

Due to the complexity of the spin-phonon system, the numerics for ground-state cooled crystals already exhaust the capabilities of classical computers. However, by studying the corresponding Heisenberg equations,  one finds $\langle\sigma_j^x(t)\rangle=\langle\sigma_j^{x}(t)\rangle_{\rm eff}(1-\epsilon)$, where the finite temperature leads to  $\epsilon\propto\sum_n\bar{N}_{n}^{\bot}\Omega_{\rm L}^2\eta_{n\bot}^2\mathcal{M}_{jn}^{\bot}\mathcal{M}_{jn}^{\bot}/\delta_{n\bot}^2,$ with $\bar{N}_{n}^{\bot}=\langle a_n^{\dagger}a_n\rangle_{\rm {th}}$. Therefore, there exists a trade-off between strong spin-spin interactions and small residual errors, namely, the larger the laser Rabi frequency is, the stronger the spin interactions become but also the bigger the residual error gets. We have found a compromise for the two, such that for tight confinement and the above parameters, this error is roughly $\epsilon\approx10^{-2}\bar{N}$, and thus ground-state cooling is not required.  Another possible source of error is the  micromotion~\cite{wineland_review}, which may cause heating, or  modify the spin-dependent force by  additional sidebands.  The heating is negligible far from the structural transition where non-linearities are small. Additionally, the effect of the  micromotion sidebands can be  neglected if $\omega_0\gg\Omega_{\rm rf}$, and $\omega_{\rm L}\approx \omega_y\ll\Omega_{\rm rf}$, where $\Omega_{\rm rf}$ is the r.f. trap frequency.

{\it Frustrated quantum Ising  model.--} To describe the essence of our quantum simulator, we   review  the main features of the  next-to-nearest neighbor quantum Ising model  ($J_1$-$J_2$-$QIM$), 
$
H=J_1\sum_j \sigma_{j}^z\sigma_{j+1}^z+J_2\sum_{j}\sigma_j^z\sigma_{j+2}^z-h\sum_j\sigma_j^x,
$
in the frustrated regime  ${\rm sign}(J_2)=-{\rm sign}(J_1)=+1$. For both  $J_2,h\to 0$, one finds  a ferromagnetic  two-fold ground-state 
$
\ket{{\rm F}}\in{\rm span}\left\{\left |^{} _ {\uparrow} {}^{\uparrow} _{} {}^{} _ {\uparrow} \cdots ^{} _ {\uparrow} {}^{\uparrow} _{} {}^{} _ {\uparrow}\right \rangle, \left |^{} _ {\downarrow} {}^{\downarrow} _{} {}^{} _ {\downarrow} \cdots ^{} _ {\downarrow} {}^{\downarrow} _{} {}^{} _ {\downarrow}\right \rangle  \right\}.
$
For $J_1,h\to0$, the ground-state is  a four-fold dimerized anti-ferromagnet 
$
\ket{{\rm AF}}\in{\rm span}\hspace{-0.2ex}\left\{\left  |^{} _ {\uparrow} {}^{\uparrow} _{} {}^{} _ {\downarrow}{}^{\downarrow} _ {} \cdots ^{} _ {\uparrow} {}^{\uparrow} _{} {}^{} _ {\downarrow}{}^{} _ {}\right \rangle\hspace{-0.4ex},\hspace{-0.3ex} \left |^{} _ {\downarrow} {}^{\downarrow} _{} {}^{} _ {\uparrow} {}^{\uparrow} _{} \cdots ^{} _ {\downarrow} {}^{\downarrow} _{} {}^{} _ {\uparrow}\right \rangle\hspace{-0.4ex},\hspace{-0.3ex}   \left |^{} _ {\downarrow} {}^{\uparrow} _{} {}^{} _ {\uparrow} {}^{\downarrow} _{} \cdots ^{} _ {\downarrow} {}^{\uparrow} _{} {}^{} _ {\uparrow}\right \rangle\hspace{-0.4ex},\hspace{-0.3ex} \left |^{} _ {\uparrow} {}^{\downarrow} _{} {}^{} _ {\downarrow} {}^{\uparrow} _{} \cdots ^{} _ {\uparrow} {}^{\downarrow} _{} {}^{} _ {\downarrow}\right \rangle\hspace{-0.4ex}\right\}\hspace{-0.2ex}. 
$
The {\bf F}-{\bf AF} phase transition  is  known to be of first order with an exponentially degenerate ground-state at $|J_1|=2J_2$,  which is a hallmark of frustration.  Making  $h\gg |J_1|,J_2$, the ground-state becomes a paramagnet with the spins aligned with the transverse field
$
\ket{{\rm P}}=\left |^{} _ {\rightarrow} {}^{\rightarrow} _{} {}^{} _ {\rightarrow}{}^{} _ {} \cdots ^{} _ {\rightarrow} {}^{\rightarrow} _{} {}^{} _ {\rightarrow}{}^{} _ {}\right \rangle  
$. The transitions between the ordered phases mentioned above and the disordered paramagnet  yields a rich phase diagram [Fig.~\ref{spin_dynamics}{\bf (c)}]. Indeed,  both numerical~\cite{anni_dmrg} and analytical~\cite{allen_bosonization} methods hint at the existence of a new phase separating the {\bf AF-P} transition. This critical phase, commonly denoted as the {\it floating phase} ({\bf FP}), exhibits  quasi-long range order in combination with an incommensurate periodic modulation of the spin-spin correlators  that depends on the couplings. We  stress that the existence of the {\bf FP} has raised some controversy~\cite{controversy_floating}, which makes a QS of the utmost interest to settle the problem.

Our QS  naturally includes the effect of longer-range interactions, which may be expected  to play an important role in the presence of frustration. Using a  Lanczos algorithm optimized for our  dipolar model~\eqref{quantum_ising}, we have diagonalized  systems of up to $24$ spins, and considered a homogeneous lattice spacing with $\sigma^{\alpha}_{N+1}=\sigma_{1}^{\alpha}$ in order to capture the  bulk properties of the ion crystal. To identify the different phases, we have computed the magnetic structure factor $S_{zz}(q)=\sum_{j,k}\langle\sigma_j^z\sigma_k^z\rangle\ee^{\ii q(j-k)}$ with $q\in[0,2\pi)$, which is  an order parameter accessible  in trapped-ion experiments (see below).  In  Fig.~\ref{spin_dynamics}{\bf (d)}, we summarize our main results, according to which  the ordered phases of the next-to-nearest neighbor model survive to the longer-range interactions. This is evidenced in Fig.~\ref{spin_dynamics}{\bf (e)}, where we show that the $S_{zz}(\pi/2)$ component of the structure factor captures   the dimerized-antiferromagnetic  and the floating phases. As one crosses the floating phase and  enters in the paramagnetic region, the two different types of decay are consistent with the {\bf AF}-{\bf FP}  Pokrovsky-Talapov and the continuous {\bf FP}-{\bf P}  Kosterlitz-Thouless phase transitions. 

Remarkably, our results show that the longer dipolar range  is crucial in the strongly-frustrated regime $|J_1|\approx2 J_2$. We find that the additional couplings split the  {\bf F-AF} critical point, and moreover we conjecture that they give rise to a new phase [Fig.~\ref{spin_dynamics}{\bf (d)}].   
The different ordering in this central region is a consequence of the competing frustration mechanisms introduced by the long range [Fig.~\ref{ladder_scheme}{\bf (d)}].  By including consecutive terms of the dipolar tail, we have checked that  the splitting of the critical points is caused by the third- and fourth-nearest neighbors couplings. By increasing these couplings at the expense of the nearest- and next-to-nearest-neighbor interactions, the size of this central region grows. Besides, its robustness is not modified by including further longer-range terms, such as fifth- and sixth-neighbor interactions.   In Fig.~\ref{spin_dynamics}{\bf (e)}, one observes that this new central region supports a different ordering, as evidenced by the existence of a modulation in the structure factor with momentum $q=\pi/4$ not present in the {\bf F}, and {\bf AF} phases. We have checked  the decay of this component  as one moves into the {\bf P} phase,  which supports this conjecture.

{\it Efficient detection methods for the phase diagram .--} Since full quantum state tomography becomes  inefficient for many-body systems, we  focus on the  measurement  of the order parameters above. One of the advantages of trapped ions  with respect to other platforms is their ability to perform high-accuracy measurements  at the single-particle level~\cite{wineland_review}.  From the spatially resolved  state-dependent fluorescence, one can infer   local observables  $m_{j,z}=\langle\sigma_j^z\rangle=\frac{1}{2}(P_j^{e}-1)$, or  two-body  correlators $C_{zz}({\bf r}_j^0,{\bf r}_{k}^0)=\langle\sigma_j^z\sigma_{k}^{z}\rangle=\frac{1}{4}[1-2(P_j^{e}+P_{k}^{e})+4P_{j,k}^{e,e}]$.
An alternative  is to measure global properties  of the  light emitted by the ion ensemble, which carries information about the spin correlations~\cite{collective_fluorescence}. The fluorescence  spectrum along a detection direction, $\hat{\bf r}$, is related to the structure factor $\mathcal{S}_{\hat{\bf r}}(\omega)\propto\sum_{ij}\ee^{\ii k{\bf \hat{r}}\cdot({\bf r}^0_i-{\bf r}^0_j)}\langle(1+\sigma_i^z)(1+\sigma_j^z)\rangle$, where $k=\omega/c$ is the wavevector of the emitted light.  Even if the ion spacing is larger than the optical wavelength, one may compensate it by placing the detector orthogonal to the crystal plane, and using a good angular resolution. Note that such  structure factors may be used to obtain a lower bound on  entanglement~\cite{ent_bound}.

{\it Conclusions and outlook.--} We have demonstrated theoretically that cold Coulomb crystals can be used as quantum simulators of paradigmatic, yet controversial, models of frustrated quantum spins. We presented a method to control the {\bf F}-{\bf AF} nature of an Ising spin-spin interaction between the ions trapped in a zig-zag structure, leading to a quantum simulator of the  $J_1$-$J_2$ quantum Ising model with  dipolar range. We have discussed the experimental feasibility of the proposed work and corroborated it by numerics  based on experimentally realistic parameters.    Analogous ideas can be transferred to different geometries in self-assembled Coulomb crystals, or surface microtraps, to realize other frustrated  models. These ideas might also be relevant to atoms confined in multimode optical cavities

{\it Acknowledgements.--} This work was supported by the EU STREP projects
HIP, PICC, AQUTE, QESSENCE, and by the Alexander von Humboldt Foundation. We thank R. Nigmatullin and D. Porras   for useful discussions.

\vspace{-4ex}

\end{document}